# Coexistence of different pinning mechanisms in Bi-2223 superconductor and its implications for using the material for high current applications

Md. Arif Ali, S. S. Banerjee*

Department of Physics, Indian Institute of Technology Kanpur, Kanpur - 208016, Uttar Pradesh, India

**Abstract**: We investigate the pinning mechanism in high-critical-current polycrystalline samples of Bi-2223 ($Bi_2Sr_2Ca_2Cu_3O_{10}$) utilized in high current applications. Using differential magneto-optical (DMO) imaging technique, we track the magnetic field penetration in the sample. Our DMO imaging studies show circular regions with an average diameter of 20 μm with dark contrast. We identify these as strong-pinning regions with a substantially higher local penetration field than the surrounding regions. A unique feature of these strong-pinning centers is that they survive upto high temperatures (near $T_c$) and produce a non-Gaussian distribution of the penetration field strength. By analysing the field dependence of the pinning force behaviour, we identify two distinct pinning mechanisms: at low temperatures, well below $T_c$, it is predominantly surface pinning mechanism and at higher temperatures near $T_c$, we see a crossover into a purely $\delta T_c$ pinning mechanism. Our studies show that surface pinning effects are related to grain alignment, grain boundary, and voids in the sample. The effect of these diminishes near $T_c$, and the strong-pinning regions here are related to local stoichiometric fluctuations. We investigate the impact of these pinning centers on the current distribution in a macroscopic Bi-2223 superconducting cylindrical tube. We map the current distribution across the macroscopic cylindrical tube using an array of hall sensors distributed around the cylinder. The map shows that the current distribution is non-uniform across the tube at high currents. The non-uniformity reveals an inhomogeneous distribution of strong-pinning centers across large length scales in superconductors used for applications.

*Email: satyajit@iitk.ac.in

A. Introduction

A useful property of a type II superconductor is its critical current density ($J_c$), which is the maximum dissipationless current density sustained by the superconductor. In a type II superconductor, the vortices are pinned by inhomogeneities in the sample. The upper bound of $J_c$ is the depairing current density ($J_0$) of a superconductor which is the current at which the kinetic energy of the Cooper pairs overcomes the condensation energy of the pairs. In real superconductors, the $J_c$ is restricted to much lower values than $J_0$. At a magnetic field $B$ applied to a type II superconductor, the $J_c$ is the bulk vortex pinning force density $F_p$ which is equal to the Lorentz force $= J_c \times B$ required to just depin the (pinned) vortices when a current density $J = J_c$ is sent across the superconductor. Vortex pinning arises from a complex interplay of different energy scales, like the vortex line tension, the inter-vortex interaction, the vortex-pin interaction, and thermal energy[1]. A lot of research effort continues to be dedicated towards searching for more efficient sources of natural pinning in superconductors as well as searching for newer ways of producing artificial pinning centres, all of which can result in high $J_c$ approaching close to $J_0$ value of a superconductor. In this regard, a lot of research focus has been in high-temperature superconductors as they possess a high superconducting transition temperature $T_c$, which makes these materials lucrative from the applications point of view as applications based on these superconductors can be operated at liquid Nitrogen temperatures. The combination of achieving high $J_c$ along with high $T_c$ is a primary challenge in the area of applied superconductivity[2,3,4,5,6,7,8,9]. Strong pinning in a superconductor can arise naturally due to chemical impurities, crystalline imperfections like vacancies, defects and dislocations in the crystal or can also be generated artificially through heavy-ion irradiation[1]. Very effective ways of enhancing pinning strength artificially are via heavy ion irradiation and chemical doping.[1,10,11,12,13]. High temperature superconductor tapes at 4.2 K operating in an applied magnetic field of 5 T have reached $J_c$ = 8 MA.cm$^{-2}$,[11] which is still a few orders of magnitude smaller than the $J_0$ of the material. Most applications employing superconductor's usually use macroscopic-sized high-temperature superconductors (HTSC's). Using HTSC's in such large quantities is only possible by using polycrystalline forms of the material. Hence from the applications point of view it is important to discuss the limitations of $J_c$ of polycrystalline HTSC materials. A fundamental limitation of $J_c$ in polycrystalline HTSC's arises from the inability of grain boundaries in the polycrystal to sustain large critical currents. Typically, it has been found in some polycrystalline HTSC materials for e.g., YBCO, that reducing the grain orientation mismatch angles to less than 4° can lead to a large increase $J_c$ [13]. In YBCO and Bi-2212, studies show that the ratio of intergrain to intragrain critical current density decreases exponentially with increasing tilt angle.[14] Another important aspect affecting $J_c$ is the effect of thermal fluctuations. For most applications using HTSC's, the operating temperature is liquid Nitrogen (77 K) or higher. Hence from the applications point of view apart from investigating polycrstalline forms of a material, it is also important to study pinning effects and understand the fundamental limitations to achieving high $J_c$ at $T$'s which are much higher than 4.2 K. Enhanced thermal wandering of vortices from their mean position at $T$ near $T_c$ of a superconductor, causes the vortices to depin. Consequently most pinning centre's become ineffective at high $T$ (near $T_c$) and consequently, the $J_c$ of most HTSC materials decrease quite rapidly at high $T$ (especially at 77 K and higher)[1]. Herein, one would like to search for strong pinning mechanism's which remain effective upto high $T$'s which are close to $T_c$. The most popular HTSC's materials used in commercial applications has been $YBa_2Cu_3O_7$ (YBCO) and $Bi_2Sr_2Ca_1Cu_2O_8$ (Bi-2212). The Bi-2223 phase ($Bi_2Sr_2Ca_2Cu_3O_{10}$) is also an important HTSC material as it has a $T_c$ = 110 K which is higher than that of YBCO and Bi-2212. With a lot of research focus being on YBCO tapes as conductors, the focus on Bi-2223 has been comparatively less. It maybe mentioned here that the process for making wires of Bi-2223 is comparatively simpler compared to manufacturing tapes of YBCO and Bi-2212. Almost all studies have been on Bi-2223 tapes[15,16,17,18,19,20,21], where the material is subjected to significant mechanical processing to be made drawn into tapes. Such processing itself can modify the intrinsic pinning properties of a superconducting material. All these studies on Bi-2223 report the pinning potential in the Bi-2223 tapes are ~ 900 – 1000 $k_B$.[15] The earlier studies on Bi-2223 tapes indicated collective pinning behaviour in the Bi-2223 tapes[1,15]. This suggests a relatively weak pinning energy scale, as collective pinning behaviour arises when the vortex-pinning interaction energies are at a comparable scale with the intervortex interaction energy scale. Studies suggested that the presence of weak links at the grain boundaries in the material limits the critical current to comparatively low values[16] compared to other popular HTSC tapes. The current distribution in these tapes was also found to be filamentary[15,16,18]. While these preliminary studies have attempted to explore study on pinning in Bi-2223 tapes, they do not provide much information on the intrinsic nature of pinning in Bi-2223 polycrystalline samples, viz., is the pinning due to critical transition temperature $\delta T_c$ or mean free path $\delta l$, pinning mechanisms[1] or a competition between both. It is not clear if any Bi-2223 can have any strong pinning which can survive upto $T$ close to $T_c$. Study of such pinning centres is important as they affect the dynamics of vortices in the superconductor and inturn affect the distribution of currents in the superconductor also. In this context it is worthwhile mentioning that pinning is known to significantly affects the dynamics of vortices in a superconductor, leading to various effects like, memory effects, instable flow regimes and history dependent effects[22,23,24,25,26,27,28,29,30,31,32,33,34].

In this paper, we have attempted to study the pinning mechanism in non-tape Bi-2223 polycrystalline material. Non-tape Bi-2223 structures are typically not subjected to mechanical moulding process which are used to make specific shapes like wires or tapes after encapsulating the powder of the material into say Ag tubes. Note that these mechanical processes can modify the pinning mechanism in samples. One of the aims of the present paper is to identify sources of strong pinning which is possible in polycrystalline form of Bi-2223. The present paper also attempts to investigate how the current distribution is affected in macroscopic structures made from the Bi-2223 polycrystals studied. The polycrystalline sample we investigate is a piece taken from a commercial high current lead made from this Bi-2223 material. The sample has a high critical current of 160 A at 77 K. By using the differential magneto-optical (DMO) imaging technique, we image the advancement of the magnetic flux penetration front while increasing the field. The technique allows us to accurately determine the average penetration field of the bulk polycrystalline Bi-2223 superconducting sample. Using the DMO technique, we explore the temperature dependence of the penetration field. We find that on approaching high $T$'s which are close to $T_c$ (only above 85 K) we see in the DMO images circular regions (we call Meissner spots) with dark contrast compared to surrounding regions. The dark magneto-optical contrast represents depleted local flux in these Meissner spots which have an average diameter of 20 μm. These spots have a local penetration field which is much higher than that in the surrounding bulk. The study suggests a broad distribution of the local penetration field. By measuring the magnetic field at which these regions disappear, we find that the local penetration field and hence the local pinning inside these circular regions is much higher than the surrounding regions. The distribution in the penetration field strength produced by these strong pinning regions which remain effective even upto close to $T_c$, is found to have a wide non-Gaussian distribution. Analysis of the bulk magnetization hysteresis loops measured at different $T$'s, shows a crossover in the pinning mechanism with different pinning operating at different $T$ regimes. At relatively lower $T$ (~ 77 K to 85 K) we see surface pinning governs the pinning properties while at higher $T$ (> 85 K) $\delta T_c$ is the dominant pinning mechanism. We explore the grain morphology in our polycrystalline sample and also do a mapping of the local chemical composition mapping. Our studies show that the grains in the polycrystalline are reasonably well oriented. The surface pinning arises from pinning effects associated with grain boundary pinning, misalignment of grains and voids present in the sample. The $\delta T_c$ pinning mechanism which survives at higher $T$, arises from local regions in the sample which exhibit stoichiometry fluctuations in the Bi-2223 sample. These regions correlate with the dark spots seen to emerge in our DMO images at high $T$ (near $T_c$). Thus at low $T$ below 85 K in Bi-2223 polycrystals we find the pinning is a complex superposition of two different mechanisms, whereas at higher $T$ only one mechanism survives. To explore the uniformity of the pinning across macroscopic length scales we use an array of Hall sensors to map the current distribution across a macroscopic cylindrical tube made from the same Bi-2223 polycrystalline material investigated. These tubes are commercial high current leads used in different applications. At currents which are close to the critical current but below it, we observe an inhomogeneous current flow on the surface of these Bi-2223 tube. We uncover patches in the sample where currents avoid flowing. These regions have a lower local critical current (lower pinning strength) compared to surrounding regions with stronger pinning. Our study shows that over long length scales the distribution of the complex pinning in Bi-2223 material isn't uniform. The study reveals a need to explore ways to ensure more uniform distribution of pinning across the sample to enable homogenous critical current distribution across Bi-2223.

B. **Characterization of the superconductor**:

For our investigation, we have chosen to study two samples of polycrystalline Bi-2223 sample with a bulk $T_c$ = 110 K. For our research; we have employed and analysed the properties of Bi-2223 used in making commercial superconducting current leads obtained from CAN superconductors, Czech Republic. This macroscopic superconducting current lead is a cylindrical tube of polycrystalline Bi-2223. The superconductor is 12 cm long tube with a uniform inner diameter of 8 mm and an outer diameter of 10 mm. We use this tube to measure its critical current using electrical transport measurements and also use the same tube to map the current density distribution across this tube as the current flows down along the length of the tube. From the lower portion of this cylindrical tube we break a few pieces of the polycrystalline material for measuring their local and average bulk magnetic properties. The local magnetic properties are measured using magneto-optical imaging studies where the sample is slightly trapezoidal in shape. The perpendicular distance between the two parallel edges of the sample is 3. 2 mm and the sample thickness is 0.2 mm. For doing bulk magnetization measurements we used a rectangular shaped sample with dimensions 2.2 mm × 2.8 mm × 0.2 mm. The Bi-2223 material were prepared by solid-state reaction of oxides and carbonates of precursor powders and then subjecting them to heat treatment procedures, details of which can be found elsewhere[35,36,37].

Bulk Magnetization (*M*) versus temperature (*T*) measurements have been done on the small piece of Bi-2223 mentioned above. The measurements were performed in the SQUID magnetometer from Cryogenic, UK. After zero-field cooling the sample down to 1.2 K, data was taken while heating the sample in a small applied field of 50 Oe. The sample shows a robust diamagnetic response at low T which rapidly diminishes as the superconductor approaches the normal state at higher *T*. The *M(T)* behaviour shows a critical transition temperature ($T_c$) of Bi - 2223 sample to be 110 K. In Fig. 1(a) the $T_c$ is located at the onset of the diamagnetic *M* response. To perform current (*I*) - Voltage (*V*) measurements we used the superconducting Bi-2223 tube, which was dipped in liquid Nitrogen. We used a four-probe geometry to do the *I-V* measurements. The measured *V* is converted in electric field, *E* using $E = V/d$, where $d = 12$ cm is the spacing between the electrical contacts on the tube. In Fig.1(b) we plot *E* vs *I*. We use a 1μV/cm criterion to determine the critical current $I_c$ which is found to be around 160 A at 77 K using 1μV/cm criterion. By fitting the *E(I)* curve to the $E = E_0 + \varepsilon(I/I_c)^n$ (red curve in Fig.1(b), where $\varepsilon = 1\mu V/cm$, we get $I_c = 158.8$ A, $n = 11$, $E_0 = 0.00214$ mV/cm is the average noise floor of our measurement. From this value and the sample dimensions mentioned here, we estimate the bulk critical current density of the sample, $J_c^{bulk} = 5.7 \times 10^6$ A.m$^{-2}$.

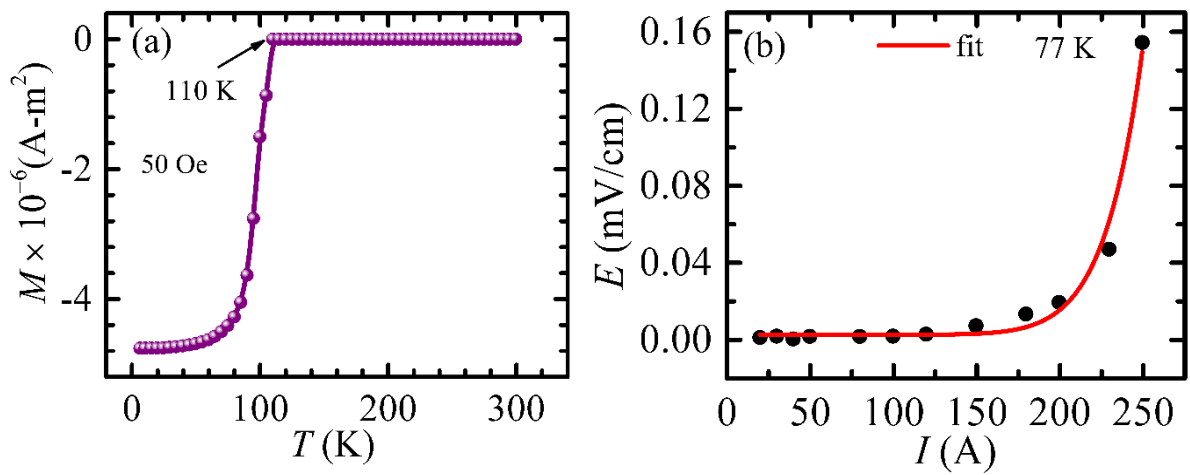

**Fig. 1**. Characterization of the sample. (a) magnetization measurement (*M*) as a function of temperature (*T*) measured in a field of 50 Oe. The black arrow shows the location of the transition temperature of 110 K. (b) Electrical transport characteristics of the sample obtained by measuring the I-V characteristics of the superconductor using four-probe measurement techniques. The data is plotted in terms of electric field (*E*) versus *I*. Using a 1μV/cm criterion, the data is fitted to an *E(I)* expression to determine the $I_c$ at 77 K.

C. **Bulk and local magnetization measurements and analysis**:

We investigate the magnetic field profiles penetrating the sample using differential magneto-optical (DMO) measurement technique[38,39]. In conventional magneto-optical (MO) imaging technique[40] a linearly polarised light is reflected from the surface of a sample on which is placed a Faraday active film with a high Verdet's constant material (YIG deposited on a GGG substrate). In this configuration, the Faraday rotation ($\theta_F$) of the reflected linearly polarized light is proportional to $B_z$, the z-component of the local magnetic field on the sample (namely, the component perpendicular to the surface of the material). By using a CCD camera (placed after an analyser in crossed position with the polarizer) the intensity distribution ($I(x,y)$) of the Faraday rotated reflected light is measured across the sample surface. By calibrating known applied fields, the $I(x,y)$ gives information of the $B_z(x,y)$ distribution across the sample (as $I \propto B_z^2$, for moderately high $B_z$ values). In our experiment, the Bi-2223 sample was cooled in nominal zero magnetic field to temperatures well below $T_C$. After stabilizing the sample at the desired $T$, a field $H$ is applied. In DMO technique a differential image is obtained at $H$ and $T$ by subtracting the average image obtained by summing 10 MO images captured at $H$ from the average of 10 image captured at a slightly incremented H value, viz., at $H+\delta H$. The number of images summed over to obtain the average image at $H$ or $H + \delta H$ can vary between 10 to 20, and it depends on improving the signal to noise ratio of the images. For obtaining our DMO images we have used $\delta H = 1$ Oe. The lower panel of images in Fig. 2 shows DMO images captured while increasing $H$ during isothermal runs, namely, (c)-(f) at 44 K. With increasing $H$ the magnetic flux begins to penetrate into the sample following the popular Bean's critical state model[41]. The schematic in Fig. 2(a) shows the typical Bean profile of the $B(r)$ distribution across the sample for different $H$ and also shown are Bean profile for the modulation in the field ($H+\delta H$) adjoining the profile at each $H$. In this schematic, the curves corresponding to $B_0$ and $B_1$ are the curves when the applied field is $H_0$ and $H_1$, where $H_0 < H_1$. The edges correspond to the sample edges from which the magnetic field begins to penetrate into the sample. With increasing $H$, the flux front begins to penetrate deeper into the sample. At each applied field, if we perform a differential measurement, then the change in the local $B$, i.e., $\delta B$ is far smaller near the penetrating front where there will be a smaller change in the vortex density compared to near the sample edges (see schematic in Fig. 2(a)). Therefore by mapping in the DMO images the regions with low DMO intensity (a region with dark contrast), i.e., $\delta I$ ($\propto \delta B_z$), we are able to trace the location of the penetrating flux front quite clearly as it advances into the sample with increasing $H$.

The DMO images in Fig. 2(c) to (f) at 44 K show the dark contour of the penetrating flux front starts from the sample boundaries and begins to move inwards into the sample with increasing $H$. In Fig. 2(b) we see the differential intensity profile $\delta I(x)$ measured along the cyan colour line drawn in Fig. 2(e). The $\delta I(x)$ profile in Fig. 2(b) clearly shows dips at the location of the dark contours. The region of the sample (unpenetrated nearly flux-free region) enclosed inside the dark penetrating boundary progressively shrinks as the $H$ increases. This process continues until the dark penetrating contour reaches the geometric sample centre. The $H$ where this happens is the penetration field $H_p$. Using this DMO procedure outlined above we can clearly measure the average $H_p$ for the bulk Bi-2223 polycrystalline sample. In conventional MO imaging, when the penetrating flux reaches the centre of the sample the brightness in the image around the centre becomes quite high, therefore it sometimes becomes difficult to accurately determine $H_p$. However, with DMO imaging, the dark contrast of the penetrating flux front allows us to determine $H_p$ with greater accuracy.

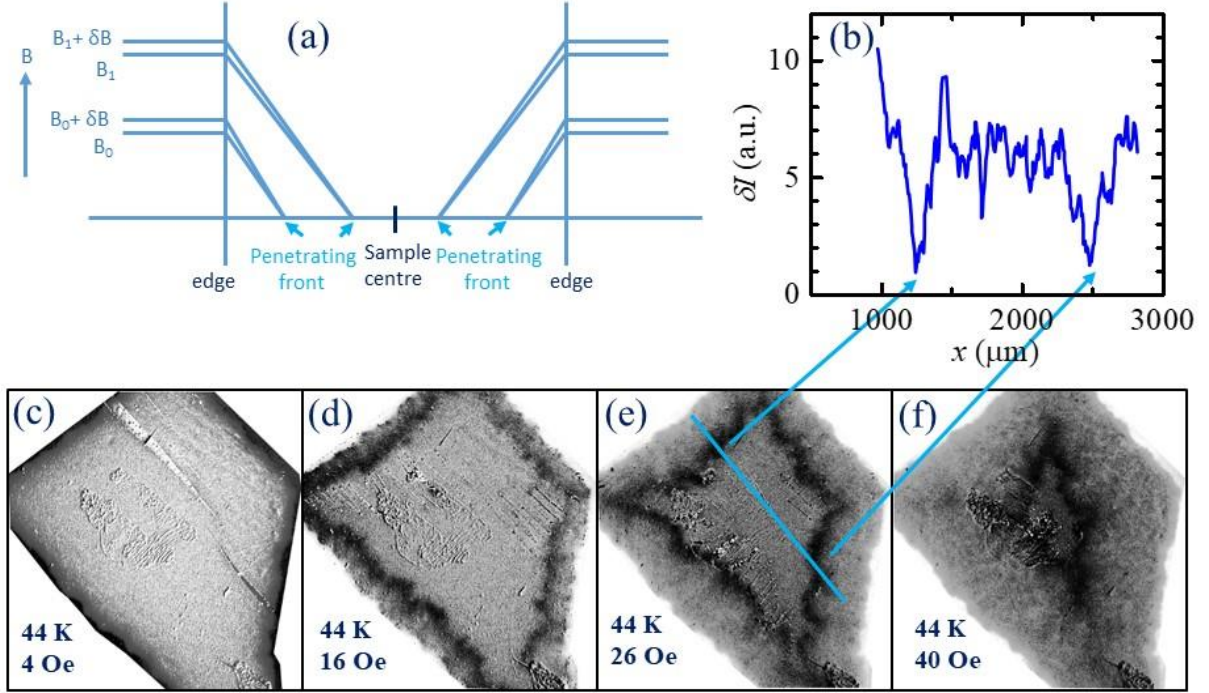

**Fig. 2**: (a) Schematic showing the Bean profile of the magnetic field ($B$) profile as the applied field is increased. The vertical lines marked edges in the schematic are the edges of the sample from which vortices start entering the sample as the field is increased. The sloping region in the profile represents the penetrating flux front. At each field $B_0$ and $B_1$ the field is modulated by a small amount $\delta B$. The difference between the profiles for $B_0$, $B_0 + \delta B$ and $B_1$, $B_1 + \delta B$ becomes smallest near the penetrating front. Panel (b) shows the $B_z$ profile along the cyan colour line drawn in (e) differential MO image at 44 K and applied field 26 Oe. (c)-(f) Differential MO (DMO) images captured at different $H$ at 44 K.

Using the above procedure, we determine $H_p$ by doing DMO at different $T$'s. In Fig. 3(a) inset we show the DMO image of penetrating contour at 60 K. In it, we consider a hypothetical square region of size 50 $\mu m \times$ 50 $\mu m$ around the geometric centre of the sample (which is marked as a red square in Fig. 3(a)). We define the $H_p$ as applied $H$ at which the dark contour in the DMO image, i.e., the region with the minimum in $\delta I$, reaches inside the red square region in Fig. 3(a). Figure 3(a) main panel shows that the spatially averaged value of $\delta I$ over the red square region at 60 K, measured at different $H$. We see that the $\delta I$ inside this region remains nearly constant and has a relatively high value for $H$ < 30 Oe. However, as $H$ approaches $H_p$ the $\delta I$ reduces to nearly zero as the dark contours begin to occupy the region inside the red square region marked in Fig. 3(a) inset. Using this produce the at 60 K the $H_p$ is found to be 35 Oe ± 1 Oe. The main panel is Fig. 3(b) shows the behaviour of $\delta I$ versus $H$ at 77 K. Here too we can identify the location of the low field $H_p$ unambiguously. As $H_p$ is directly proportional to $J_c$ of the superconductor, hence as $J_c$ reduces monotonically with increase in $T$, we observe in the inset of Fig. 3(b) a similar monotonic decrease of $H_p$ with $T$.

At 99 K the penetration field in the bulk of the sample is quite low. Even at 1 Oe (which is the resolution of our applied field) we find that this flux has already penetrated the bulk of the sample. We approximate bulk penetration field at 99 K as ~ 1 Oe. At 99 K the bulk of Bi-2223 has become reversible and bulk $J_c$ of the sample is quite low. While obtaining DMO images at 99 K which is close to the $T_c$ of the sample, most of the image looks bright. As the sample is reversible therefore the change in the local magnetic field $\delta B_z$ ~ 1 G follows the modulation of 1 Oe of the applied field which is done to obtain the DMO images. Due this change in local field the local DMO intensity in the nearly reversible portions of the sample will be higher compared to the regions where pinning still dominates. In our DMO images we observe the presence of dark spots in the images, while the differential intensity in the vicinity of the spots is bright. We see this type of feature appears in the DMO images only above 85 K (see supplementary information of images at 87 K). At lower $T$ (below 75 K) we do not see such features, for e.g. see DMO images at 44 K in Fig. 2(c). In Fig. 3(c) we see these spots are present in the DMO images at different $H$. However, with increase in $H$ these spots gradually disappear (at high $H$). At 100 Oe we almost do not see any spots in the images in Fig.3(c). Henceforth we refer to these spots as Meissner spots (MS). These diameter of the dark spots in the images varies between 10 to 30 μm, with an average diameter of about 20 μm. The dark contrast in the MS regions is due to a depleted flux density compared to the surrounding regions of the sample. This clearly suggests that the MS regions have a higher local penetration field compared to an average bulk penetration field of 1 G at 99 K. We call the spots as Meissner spots because of they behave like Meissner like regions as the flux cannot penetrate these spot like regions whereas in the regions outside these spots the flux has already penetrated. These MS regions correspond to strong pinning regions in the sample as they possess a much higher local penetration field. To estimate the spread in the distribution of the local penetration field caused by these strong pinning centres, in the inset of Fig.3(d) we count the number of MS, $N_{MS}$, where $\delta I$ in the dark regions is less than half of value of $\delta I$ in the surrounding regions, as a function of $H$. Such spots have been circled in red in the inset figure. We determine the behaviour of $N_{MS}$ vs $H$ at 99 K in regions of the sample with high density of MS. The behaviour of $N_{MS}$ vs $H$ is shown in Fig.3(d) main panel. We clearly see a monotonic decrease of $N_{MS}$ as a function of $H$ until by about 40 Oe we are unable to determine sufficient number of MS in the region of sample compared to the number in that region at low fields. Thus the distribution of the strong pinning regions whose pinning can survive upto $T$'s close to $T_c$ gives rise to a broad non-Gaussian distribution of the penetration field in the sample. The non - Gaussian distribution of the penetration field strength across these pinning centres we believe is related to the material processing conditions[35,36,37] via which the Bi-2223 sample was synthesized. To understand the MS features, we perform magnetization ($M$) - $H$ measurements at different $T$.

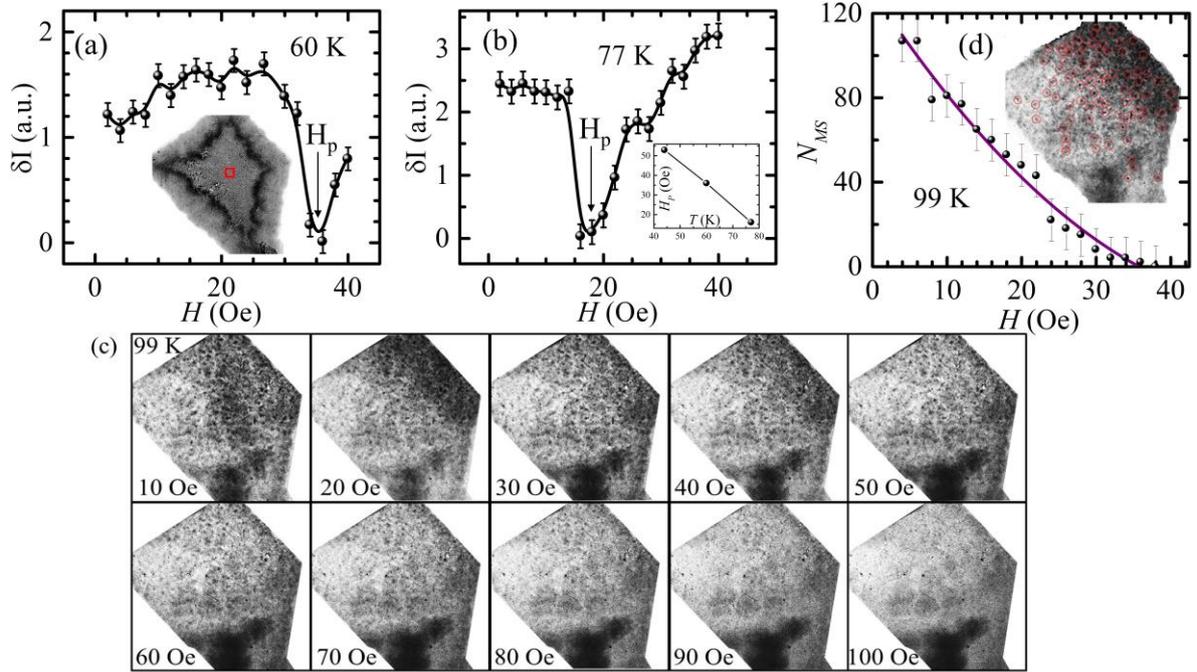

**Fig. 3**: (a)-(b) Temperature Dependence of penetration field at 60 K and 77 K respectively. (c) MH measurement of Bi-2223 sample at 99K, which is around the transition temperature of the sample 110 K. (d) The number of Meissner spots as a function of the applied field at 99 K.

Figure 4(a) shows the open $M$-$H$ hysteresis loop at 5 K showing the strong irreversibility due to vortex pinning in the superconductor. At 77 K the $M$-$H$ hysteresis loop width has reduced and the five quadrant loop in inset of Fig. 4(b) appears to be reversible and diamagnetic (in the virgin and forward leg of the *M-H* loop). As the main Fig. 4(b) shows, the loop is reversible only at high *H* beyond 0.25 T. Below 0.2 T the loop is irreversible. The Fig. 4(c) inset shows the M-H loop recorded at 100 K which is measured upto ± 6 T. The inset shows the *M-H* loops appears to be reversible, however upon measuring the *M-H* at lower magnetic fields (main panel of Fig. 4(c)) we observe signatures of hysteresis persisting at low *H* upto ~ 0 to ☐0.1 T. From the *M-H* hysteresis loops, we determine the loop width ΔM, which is the difference between the forward and reverse *M*. From the M-H loop at 77 K, using ΔM value, we estimate the bulk critical current density at nominally zero-field using, $J_c^{bulk} = 20\,\Delta M/[a(1 - a/3b)]^{42}$ where $a = 2.2$ mm and $b = 2.8$ mm are the crystal dimensions perpendicular to external field $B_a$. Using these for the M-H loop at 77 K we estimated $J_c^{bulk} = 2 \times 10^6$ A.m$^{-2}$ in nominally zero field compares well with the value of $5.7 \times 10^6$ A.m$^{-2}$ obtained from the *I-V* measurements in Fig. 1. In Fig.4(c) inset we plot the behaviour of the pinning force density ($f_p = J_c \times H$) of the Bi-2223 sample on a log linear scale. At 5 K we observe the $f_p$ doesn't reach a maximum in the field regime of interest. At 77 K there is a peak in $f_p(H)$ (marked by black arrowhead) which shifts to a higher *H* value at 100 K.

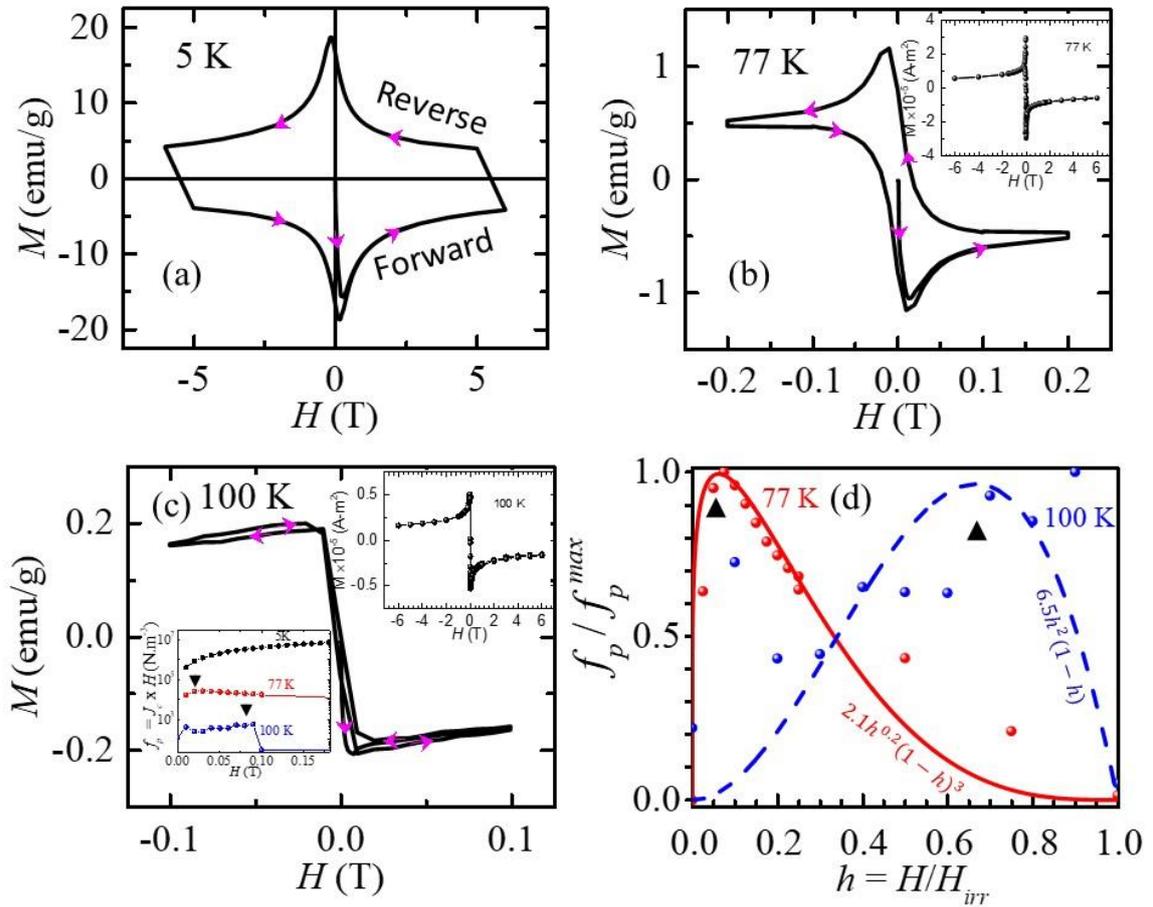

**Fig. 4**: (a), (b) and (c) show the five quadrant *M* vs *H* hysteresis loop of the sample recorded at 5 K, 77 K and 100 K in the Quantum Design SQUID magnetometer. The five quadrant loop corresponds to (i) virgin *M-H* curve for *H* being increased from nominal 0 field to a target maximum field, (ii) reverse *M-H* curve for *H* being decreased from target maximum field to a nominal 0 field, (iii) the negative field version (i) and (iv) is the negative field version of (ii) and (v) is repeat of (i). The arrows on the M-H loop represent the direction of field change for this five quadrant measurement of *M-H* loop. The inset of Fig.(b) and upper inset of (c) shows the full hysteresis loop. Lower inset of Fig.(c) shows the pinning force per unit volume ($f_p$) as a function of *H* for 5 K, 77 K and 100 K. The black arrows heads represent peak in $f_p(H)$ behaviour. (d) is the scaled pinning force plot of $f_p/f_p^{max}$ versus $h=H/H_{irr}$ for 77 K and 100 K data. The data is fit to different models (see text for details).

To analyse these features, we normalize the $f_p$ data by its maximum value at the peak and $H$ is normalized by $H_{irr}(T)$, which is the irreversibility field at the temperature $T$. Here we would like to mention that while one typically normalizes $H$ using the upper critical field $H_{c2}$, however in high $T_c$ superconductors the value of $H_{c2}$ is very high and difficult to experimentally determine. It has been shown that $f_p$ data can be scaled and analysed by normalizing $H$ using $H_{irr}(T)$ [43,44]. In Fig.4(d) for 77 K and 100 K, we plot $f_p/f_p^{max}$ versus $h = H/H_{irr}$, where $H_{irr}(77 K) = 0.4$ T and $H_{irr}(100 K) = 0.1$ T. We see that $\frac{f_p}{f_p^{max}} = 2.1h^{0.2}(1-h)^3$ fits with the data for 77 K while the data for 100 K fits to $\frac{f_p}{f_p^{max}} = 6.5h^2(1-h)^1$. The pinning force data for our Bi-2223 sample is analysed using the Dew Hughes scaling form [45], $\frac{f_p}{f_p^{max}} = Ah^p(1-h)^q$, where $A$ is a constant and $p$ and $q$ are parameters whose value determines the pinning mechanism. As per the Dew Hughes [43,46] the value of $p = 0.5$ and $q = 2$ represents normal surface pinning mechanism while $p = 2$ and $q = 1$ represents $\delta T_c$ core pinning [1]. From the fit, it is clear that at lower $T$ of 77 K the fitting form is close to $p = 0.5$ and $q = 2$ which suggests a dominance of surface pinning while for higher $T = 100$ K the pinning mechanism is dominated by $\delta T_c$ pinning. Surface pinning describes pinning centre governed by microstructural and geometry aspect of the surface. Studies in polycrystalline samples of superconductors like PbIn and Nb have shown that surface roughness and defects on the surface control the critical current in these samples[47,48,49]. While these studies showed that surface pinning affects vortex motion at currents above the bulk critical current of the superconductor, however it was not clear from these studies if this is the dominant pinning mechanism at all temperatures (even near $T_c$) and does it also operate in Bi-2223 polycrystals. From Fig. 4(d) we see that with change in $T$ the pinning mechanism completely changes from surface pinning to $\delta T_c$ pinning. The dark MS regions with a typical diameter of ~ 10 μm we see appearing in the DMO images at 99 K in Fig. 3, are the $\delta T_c$ pinning sites. Due to strong pinning the local $B$ inside these MS regions doesn't change with modulation in $H$, i.e., the magnetic flux doesn't penetrate these regions and therefore $\delta B$ ($\delta I$) is small inside the MS regions. Outside the MS regions, the superconductor is reversible at 99 K and in these regions the magnetic flux freely penetrates. Due to this at 99 K the reversible regions have $\delta B \approx \delta H$ which is higher than that in the MS regions. At 99 K (Fig. 3) with the increase in $H$ as the MS regions locally become reversible, the contrast between the $\delta I$ value inside the MS regions and outside it lessens and hence these regions become faint and disappear from view in the DMO images as the $H$ is increased (see Fig. 3(c)). At lower $T$ pinning is primarily governed by surface pinning effects compared to $\delta T_c$ pinning. Hence surface pinning essentially determines the bulk penetration fields in the sample and hence the bulk $J_c$ values. However, at higher $T$ (near $T_c$) while most of the sample has become reversible finite pinning effects are still retained in the sample locally. In this $T$ regime compared to surface pinning effects $\delta T_c$ pinning effects become stronger. We would like to emphasize here that even at 77 K, the $p$ and $q$ parameters we have obtained do not represent only surface pinning as their values are different from those associated with pure surface pinning. Hence we feel at 77 K which is the optimal temperature at for applications based on Bi-2223, the pinning is a combination of surface and $\delta T_c$ pinning effects, where the surface pinning is likely to play a dominant role. This also suggests that for improving pinning at 77 K regime it is more effective to do modifications of the surface in order to affect the pinning properties.

D. **Sources of pinning in Bi-2223 polycrystals**:

In Fig. 5 we study the microstructure of the Bi- 2223 sample. In Fig. 5(a) and (b) the Field Emission Scanning Electron Microscope (FESEM) image of the polycrystalline Bi-2223 sample shows several grains with a very wide distribution in the grain size. The typical maximum length of grain was found to be around 20 μm. There is a large amount of voids (dark regions) and cracks in the investigated sample area. Energy Dispersive X-ray (EDX) analysis in Fig. 5(c) shows a uniform concentration of Bi, Sr, Ca, Cu, and O elements across the sample. Uniformity of colour in each image of Fig. 5(c) represent the concentration of a particular element uniform over the region of the sample shown in the image. Figure 5(d) shows the crystallographic orientation mapping of the grains on the surface of the sample using Electron Backscatter Diffraction (EBSD) technique. The colour coded stereographic triangle shows the different orientation mapping in the direction normal to the sample surface (i.e., in [001]-direction).

Form the software shaded EBSD image in Fig. 5(d) we see that the sample contains a significant number of grains which are well aligned or have similar orientation within the grains (see the similar greenish coloured grains in Fig. 5(d)). A careful study will show that within the green coloured grains there are regions with a lighter shade of green which suggests the presence of slight misalignment of grains. The image in Fig.5(d) also shows the presence of grains with distinctly different crystallographic orientations (regions with colours other than green). Recall here that low misalignment of grains is important for enhancing the inter-grain critical current density of polycrystalline superconductors[13]. The pinning associated with grain boundaries and grain alignment constitutes surface pinning effects in our Bi-2223 polycrystal. In the images of Fig. 5(c) while it appears that the average chemical composition is almost uniform, there are clear local variations which are visible. There are regions where the concentration of certain compounds is either high (brighter contrast compared to surrounding regions) or low (darker contrast compared to surrounding regions). For example, in Fig. 5(c), some local bright spots in the sample appear to have higher Bi and Pb content, and in these regions, the Sr and Ca and Cu content is lower (see the region inside the dashed black circle). These local off stoichiometry regions in Fig. 5(c) coincides well with the features seen in our DMO images at $T$ near $T_c$, viz., the dark spots which correspond to higher local penetration field in Fig. 3(c). At these $T$ our analysis of the pinning force in Fig. 4(d) reveals the $\delta T_c$ pinning mechanism. Thus from the above suggests that surface pinning effects (found in Fig. 4(d), red curve) most likely is to the existence of misaligned grain, pinning from grain boundaries, voids present in the sample and the localized regions with stoichiometry fluctuations. While at higher $T$ near $T_c$, the pinning effects related to grain boundaries diminish and the $\delta T_c$ pinning mechanism is produced by local off stoichiometry fluctuations. Recall our DMO images in Fig. 3(c) have already shown that while most of the superconductor is driven into a reversible regime due to enhanced thermal fluctuations, the pinning in the MS regions is retained. We would like to mention that in Bi-2223 it appears that local off stoichiometry fluctuations seem like a more effective was to generate strong pinning in the sample as the effect of these pinning centres remains effective even in the enhanced thermal fluctuations regime near $T_c$.

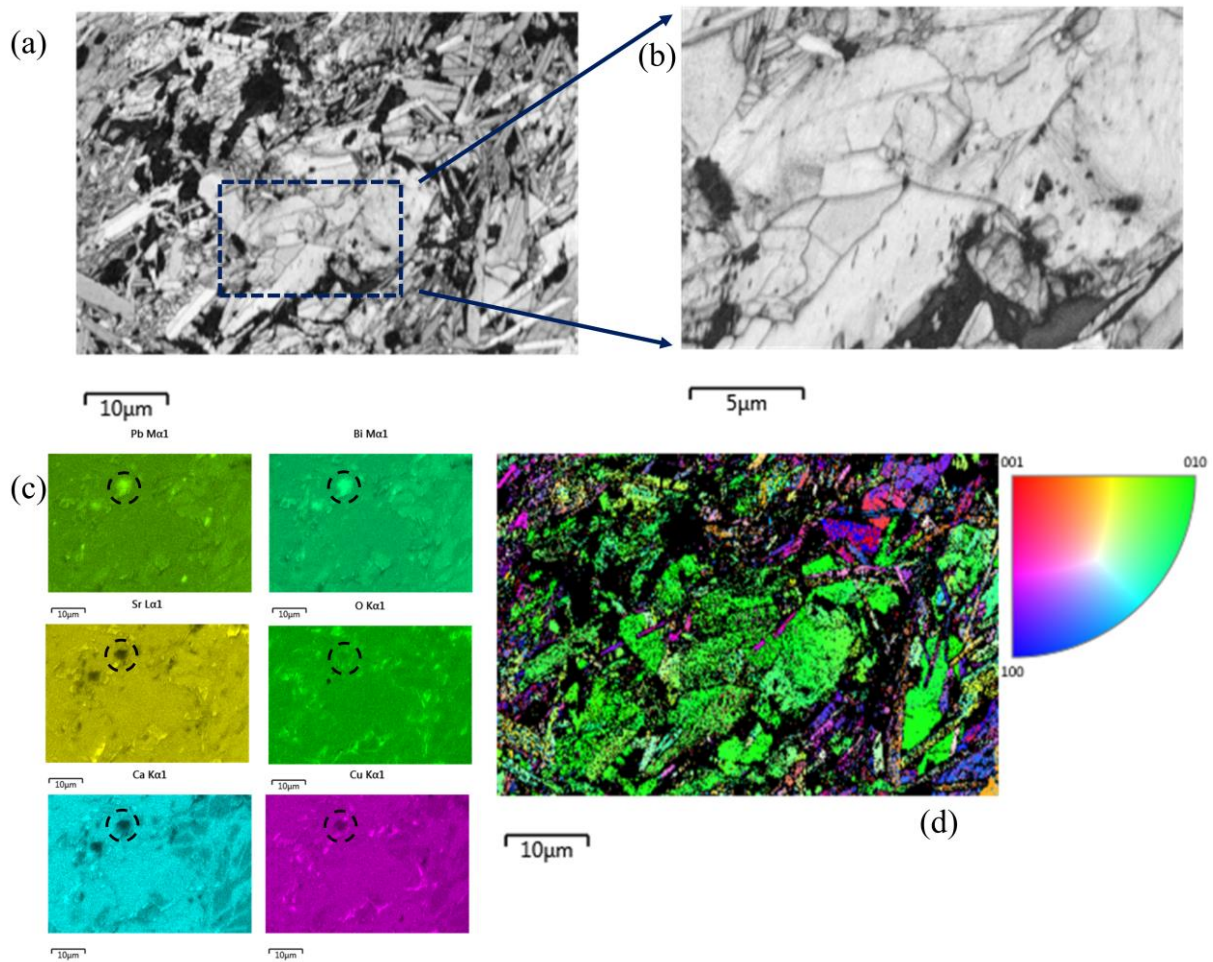

**Fig. 5**. Microstructures of Bi-2223 bulk investigated by FESEM and EBSD. (a) FESEM image of polycrystalline bulk Bi-2223 HTSC. The image shows grains with a size distribution. The maximum length of a typical grain is around 20 μm. (b) magnified region of the image in (a) with dotted square. (c) EDX of bulk Bi-2223 which show uniformity of the concentration of Pb, Bi, Sr, O, Ca, and Cu across the sample. (d) EBSD image of the Bi-2223 indicates differently oriented grains with changing boundaries.

E. **Distribution of currents across a macroscopic Bi-2223 tube**:

We investigate the behaviour of current distribution in the high current regime on the surface a Bi-2223 macroscopic superconducting tube. The superconductor we investigate is the tube of Bi-2223 which is commercially used as HTSC high current leads. A picture of the superconducting tube with Copper leads is shown in Fig. 6(a). Several groups have used Hall sensors in the past[50,51,52,53,54] to measure the local magnetization and current distribution across different superconductors. Most of these have been on small sized sample. To measure the current distribution across the macroscopic superconducting tube we use an array of 13 - AlGaAs Hall sensors connected in series. These sensors are distributed around the circumference of the tube. A schematic diagram of the position of the sensors distributed around the superconducting tube is shown in Fig. 6(a). All the Hall sensors are located approximately 5 mm above the surface of the tube. Each Hall sensor has an active area of ~ 500 μm × 100 μm with overall dimensions of the sensor being 4 mm × 5 mm × 1 mm. The sensors are oriented so that the magnetic field generated by current flowing down to the length of the tube (direction of current flow shown in Fig. 6(a)) crosses the active area of Hall sensors perpendicularly (See the dashed lines in Fig.6(a)). We use the calibration constant of each sensor to convert the Hall Voltage ($V_H(\theta, h)$) measured by each sensor into a magnetic field value ($B(\theta, h)$), where $h$ is the height of the active area of the hall sensor from the base of the tube. The calibration constant of one of the Hall sensors is ~ 90 mV/Tesla. We use a calibration procedure to convert the measured $B(\theta, h)$ into the average current density ($J_{sc}(\theta, h)$) in the region of the superconductor below the Hall sensor's location where the $B(\theta, h)$ has been measured. Briefly, the calibration procedure we use is the following: by using a Hall sensor positioned at 5 mm above the surface of a uniform cross-section copper conductor carrying a well-known current density, we calibrate the magnetic field measured by the Hall sensor versus the current density in the conductor. After this measurement, we rotate the hall sensor by $\pm\delta\theta$ about its location at ($\theta, h$) keeping $h$ fixed. We estimate the maximum $\delta\theta$ (~ 1 *radian*) variation so that $V_H(\theta, h)$ remains constant. From this we estimate that each Hall sensor measures an average current density in a patch of 5 mm × 5 mm on the surface of the superconducting tube. For more details on the calibration procedure and setup see ref [55]. For our setup the current density is measured with a resolution of 0.01 A.mm$^{-2}$. The entire superconducting tube with the Hall sensors array is dipped in liquid Nitrogen and the sensors, wiring and superconductor are all maintained at a uniform temperature of 77 K. From the $J_{sc}(\theta, h)$ measured by the array of 13 Hall sensors we construct the contour surface plot (a map) of the $J_{sc}(\theta, h)$ distributed around the superconducting tube (to make the contour surface plot we assume that the $J_{sc}$ varies continuously between two neighbouring $J_{sc}(\theta, h)$ points measured). The map of the current density around the tube is shown in Fig. 6(b) – 6(e) for different values of $I$ sent down the superconducting tube.

The $J_{sc}(\theta, h)$ image in Fig. 6(b) to (e) shows patches of low and high current density regions distributed on the surface of the Bi-2223 tube. Recall from our analysis of the average bulk $I_c$ of the Bi-2223 material used to make the tube was 160 A. We see at $I$ = 50 A (Fig.6(b)) and 100 A (Fig.6(c)) (< bulk $I_c$) the reddish patches are regions with relatively higher current density compared to the surrounding regions with greenish and bluish patches. As the current is increased to 150 A (Fig. 6(d)) which is close to $I_c$, we see a redistribution of the low and high current density patches. We see the red regions in the earlier figure of the $J_{SC}$ map turns into a low $J_{SC}$ (blue) regions. The reduction of $J_{SC}$ in these blue regions suggests that the local $I_c$ of these regions is lower than the surrounding regions. We understand this behaviour as follows: Due to the current of 150 A which exceeds the local $I_c$ in these blue regions, the trapped vortices in the blue regions of the superconductor begin to move under the influence of the

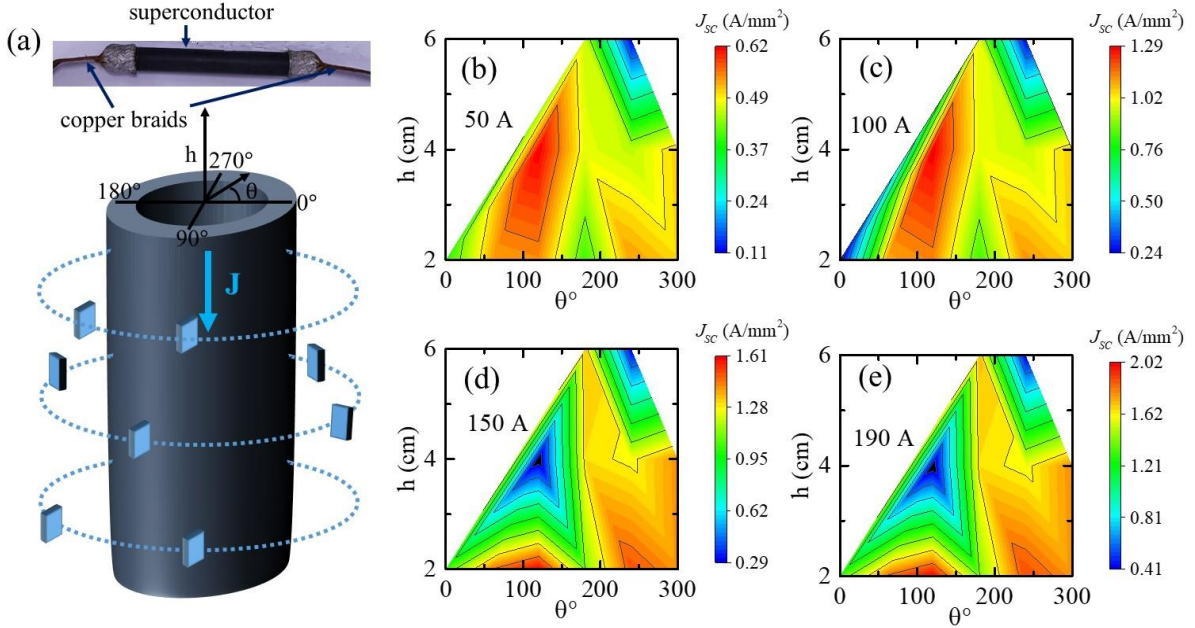

Fig 6

**Fig. 6**: (a) (above) Image of the superconductor used for current density mapping. (Below) Schematic of how Hall sensors are placed around the superconductor. Out of 13 Hall sensors which are not seen are on the other side of the superconductor. (b)-(e) Vortex redistribution after current going through the superconductor. Mapping of current density landscape over the surface of superconductor for (b) 50 A, (c) 100 A, (d) 150 A, (e) 190 A.

Lorentz force (per unit length of the region, $\vec{F_L} = \vec{I} \times \vec{B}$) and the motion generates dissipation in this region. Due to the vortex motion generating higher local dissipation inside the blue patches which is higher than the surrounding regions, the effective local resistance inside the blue patches becomes higher than the surrounding regions at these high currents which have exceeded the local critical current inside the blue regions. Consequently, the current diverts away from these high resistance blue patches in the current density map and the current redistributes itself in the surrounding regions across the tube. The map in Fig. 6(d) shows that the regions surrounding the blue regions in Fig. 6(d) have a higher critical current as the colour in these regions remain nearly uniform with change in $I$. Infact as the map at 190 A (Fig.6(e)) shows the contours of the map do not change significantly between 150 A and 190 A, this suggests the regions surrounding the blue regions in the map have local $I_c$ which is greater than 160 A. At higher $I$ = 190 A, the map shows that while the $J_{SC}$ in the blue region increases compared to that at 150 A, however, these local patches still carry a lower current density compared to the surrounding regions. This blue patches at 150 A turns red at lower $I$ = 50 A or 100 A as these $I$ values fall below the local critical current in this patch. Consequently, due to low dissipation at this $I$ more current channels into this region and it appears as a red patch at lower $I$ of 50 A and 100 A. Note that the distribution of current is controlled by the dissipation generated locally which inturn is due to motion of the vortices trapped in the superconducting tube. These vortices are not produced due to any externally applied field in our experiment. We believe they are most likely generated due to the earth's magnetic field and due to the stray field from the high current carrying leads feeding current into the superconductor (we estimate the strength of this stray field is of the order of few tens of mT at 100 A near the equatorial plane of the tube). If the net field ($B$) experienced by the superconductor ~ 50 mT then the average intervortex spacing is ~ $\sqrt{\Phi_0/B}$, ~ 0.2 μm where $\Phi_0$ = magnetic flux quantum = 2.07 × $10^{-15}$ T-m². Using our Hall sensor array setup we clearly see that there is a broad spread in local $I_c$ values across the macroscopic superconducting tube of Bi–2223. The $I$-$V$ and bulk magnetization measurements only provide an estimate of the average value of $I_c$ ~ 160 A. Thus the blue regions have suppressed $J_{SC}$ that are actually regions having higher dissipation. From these measurements, we see a spread in the local $I_c$ values leads to the creation of significant-sized macroscopic regions of size in the range of 1 cm × 1 cm. We have already discussed that at 77 K there is a combination of different types of pinning (surface pinning and $\delta T_c$ pinning) which determine the net $I_c$ of the superconductor. Out of the two different pinning mechanisms operating at different $T$, the surface pinning effects seem to be slightly weaker in strength and hence more temperature dependent pinning feature compared to $\delta T_c$ pinning which is seen to survive upto $T$ very close to $T_c$, which seems to be a more stronger pinning mechanism in the Bi-2223 sample. In Fig.3(d) we have already shown that at high $T$ the $\delta T_c$ pinning leads to a non-Gaussian distribution of penetration fields. The fact that our local $J_{SC}$ map shows the presence of relatively large sized regions with lower $I_c$ compared to neighbouring regions suggests that in macroscopic samples the distribution of pinning across the sample is not spatially uniform. In a macroscopic superconducting samples which are typically used in high current applications, our study suggests that there exist patches with relatively weaker pinning compared to other regions with stronger pinning. As our study shows that at low $T$ the surface pinning is the most dominant mechanism which arises from contributions related to grain boundary pinning, alignment of grains, defects on surface. There is also a subdominant contribution from $\delta T_c$ pinning due to local stoichiometric fluctuations. However, at higher $T$ (> 85 K) out of these different contributions only the stoichiometric fluctuations seem to be driving a $\delta T_c$ pinning mechanism. As this pinning $\delta T_c$ mechanism continues to remain effective even in the high thermal fluctuation regime close to $T_c$, we believe this pinning source is a source of strong pinning and it has a weaker $T$ dependence compared to the surface pinning effects in Bi-2223. Non uniformity in pinning strength across macroscopic length scales could be due to fluctuations in the variation of the density of strong $\delta T_c$ pinning centres across the sample. The non-uniform pinning is also due to changing microscopic surface morphology along the length of the long superconducting tube. These fluctuations in distribution of pinning strength across the sample leads to the low and high pinning region distribution found in our current map. The regions with local lower pinning strength in a macroscopic superconducting sample would lead to creation of patches with local high dissipation at high currents. These would nucleate thermal instabilities like hot spot generation from these regions. These thermal instabilities and quickly drive a superconductor to normal and at high current such a process would have an adverse impact on the performance of the superconductors in high current applications, like in superconducting fault current limiters, high field superconducting magnets. Hence it is important to find ways to obtain ensure more uniform distribution of pinning across macroscopic polycrystalline superconductors and to processing routes to enhance the density of the stronger $\delta T_c$ pinning across the Bi-2223 sample.

**Conclusion**: We find the presence of a complex pinning mechanism in Bi-2223 polycrystal. We uncover the presence of unusually strong pinning centres in Bi-2223 polycrystal, whose pinning effects survive upto $T$ close to $T_c$. At low $T$ we find pinning in Bi-2223 polycrystal is dominated by a conventional surface pinning mechanism with a subdominant contribution from $\delta T_c$ pinning mechanism. However, at higher $T$ only the $\delta T_c$ pinning mechanism dominates and the pinning is associated with regions in the sample with local stoichiometric fluctuations. While our study helps to identify a source of strong pinning in Bi-2223 polycrystals, more studies are needed to identify synthesis routes to reliably generate a uniform density of these strong pinning centres across the length of large Bi-2223 based superconductors for high current application.

**Acknowledgements**: SSB would like to acknowledge discussions with CAN superconductors. SSB also acknowledges funding support from Department of Science and Technology (Imprint II and AMT programs), Government of India and also IIT Kanpur.

# Supplementary Information

# Coexistence of different pinning mechanisms in Bi-2223 superconductor and its implications for using the material for high current applications

Md. Arif Ali, S. S. Banerjee*

Department of Physics, Indian Institute of Technology Kanpur, Kanpur - 208016, Uttar Pradesh, India

The figure below shows differential magneto-optical images at 87 K for 2 Oe. Meissner spots start appearing around this temperature.

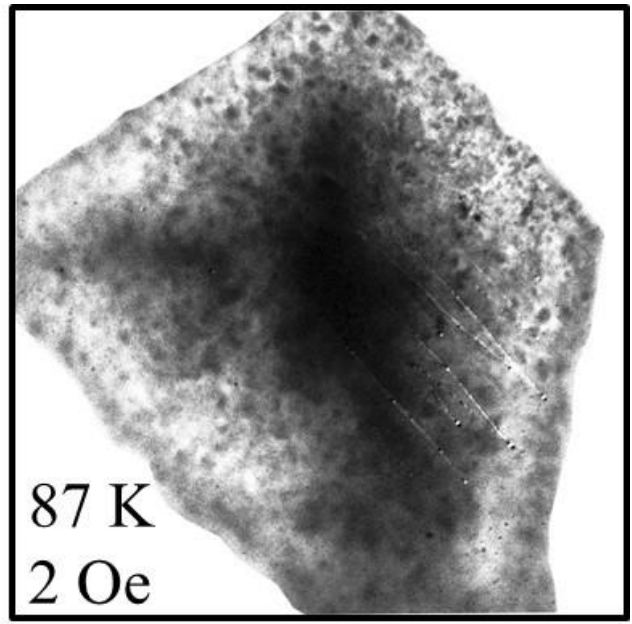

Fig. 1: Meissner spots at 87 K at 2 Oe.